# Photoluminescence from Silicon nanoparticles embedded in ammonium silicon hexafluoride

Seref Kalem[1], Peter Werner[2], Vadim Talalaev[2], Michael Becker[2], Örjan Arthursson[3] and Nikolai Zakharov[2]

[1]UEKAE, National Research Institute of Electronics and Cryptology, Gebze 41470 Kocaeli, TURKEY

[2]Max-Planck-Institute, Department of Experimental Physics, Halle(Saale), Germany

[3]Microtechnology and Nanosciences Dept., Chalmers University of Technology, Göteborg, Sweden

*s.kalem@uekae.tubitak.gov.tr*

**Abstract:**

Silicon (Si) nanoparticles (NPs) were synthesized by transforming Si wafer surface to ammonium silicon hexafluoride (ASH) or $(NH_4)_2SiF_6$ under acid vapor treatment. Si-NPs are embedded within the polycrystalline (ASH) layer formed on the Si surface exhibit a strong green-orange photoluminescence (PL). Difference measurements revealed a major double component spectra consisting of a broad band associated with the ASH-Si wafer interfacial porous oxide layer and a high energy band attributable to Si-NPs embedded in the ASH. The origin of the latter emission can be explained in terms of quantum/spatial confinement effects probably mediated by oxygen related defects in or around Si-NPs. Although Si-NPs are derived from the interface they are much smaller in size than those embedded within the interfacial porous oxide layer ($SiO_x$, $1 < \mathbf{x} < 2$). Transmission electron microscopy (TEM) combined with Raman scattering and Fourier transformed infrared (FTIR) analysis confirmed the presence of Si-NPs and Si-O bondings pointing to the role of oxygen related defects. The presence of oxygen of up to 4.5 at.% in the $(NH_4)_2SiF_6$ layer was confirmed by energy dispersive spectroscopy (EDS) analysis.

1. **Introduction**

Transformation of Si to Ammonium Silicon Hexafluoride $(NH_4)_2SiF_6$ or ASH was based on earlier gas or vapor phase processing experiments with Si wafer using vapors of $HF:HNO_3$ chemical mixture [1]. The same method has also been applied to a controlled thinning of Si nanopillars [2] and extended to the transformation of Germanium to Germanates [3]. The microscopic physical properties of the resultant material have mainly been investigated from an optical and structural point of view [3-8]. It has been shown that these fluoride or germanate layers exhibited PL features in the visible region but the full account of the origin of PL is under debate [1, 3, 5, 9]. A possible origin and the mechanism of the PL in ASH were speculated taking into account excitons, which were trapped at the energy levels of the $SiO_x$ surrounding Si nanocrystallites [9]. It was reported that defects at Si/SiOx interfaces were also responsible for the PL [9']. There has been further claim attributing the radiative recombination to free excitons in particles larger than 3 nm [9'']. But, the contribution from the ASH-Si interface is underestimated. Actually, the presence and the role of the interface layer (porous $SiO_x$) have already been described and its luminescent emission properties were investigated [1, 4, 7]. However, the likely incorporation of luminescent defect centers into the fluoride layer during the transformation process and the microscopic origin of the PL remain to be speculative and thus it needs to be explained in further detail.

The role of defects and quantum confinement in nano-crystalline (NC) Si has been demonstrated experimentally [10-10'] and a comprehensive description was given by Gösele [11]. Both of the effects can be competing with each other in determining PL properties of Si NC's. For example, an unterminated dangling bond can activate a defect mediated emission. However, terminated or passivated defect can be in favor of a quantum confinement (QC) emission. Thus, the excitation can take place in NC but the recombination can follow one of the two pathways. It was argued that there are two major defects in Si-SiOx system: iterface defects with states close to the conduction band and mid-gap defect states close to the valence band [10]. These defects can quench the PL from Si NC's in the 1.4-2.2 eV range, irrespective of whether the source is QC or interface states [12]. On the other hand, the intrinsic and extrinsic nature of PL emission from Si NC's in porous Si has been investigated in terms of hot carrier recombination at defects in Si-SiOx interface [13, 14]. The PL from Si nanocrystals embedded in SiOx has been extensively studied [15-22] . These findings emphasize the importance of a confinement and defect mediated emissions in Si-SiOx systems.

In this paper, we provide a direct evidence for the presence of the crystalline ASH clusters decorated by Si nanoparticles. The work shows that there are two main components of the visible light emission which are related to : i) Si particles in a porous oxide layer at the ASH-Silicon interface; ii) smaller Si nanoparticles embedded within the ASH host matrix.

2. **Experimental**

The ASH samples were prepared in a Teflon cell by exposing Si wafers (p-type having a resistivity in the range of 10.5-19.5 Ω-cm and <111> crystal orientation) to a vapor of $HF:HNO_3:H_2O$ (70:25:15) chemical solutions (the details of the experimental setup were described elsewhere [1] ). No initial surface cleaning was required for the Si wafers but, just before adding the water, the HF (%48) : $HNO_3$ (%65) solution was primed for about 10 second using a small piece of p-Si wafer. The wafer and the Teflon cell were kept at 300 K.

The following techniques were used for the experimental investigation: TEM (JEM 4010), scanning electron microscopy (SEM) and energy dispersive spectroscopy (SEM, JEOL-JSM-6335S), Fourier transformed infrared (FTIR) analysis, Raman light scattering and PL measurements to determine structural and optical properties. Photoluminescence was excited by HeCd laser of 23 mW at 325 nm (3.81 eV) and the signal was detected by a liquid nitrogen cooled charge coupled device (CCD). For Raman, the 488 nm line of an Ar+ laser at backscattering geometry was used.

Concerning the microstructure of the ASH, it is a polycrystalline layer which is composed of columnar structure as it was evidenced from the cross-sectional SEM and x-ray analysis [1, 7]. Some of the measurements were performed on free standing ASH layer, that is the layer which was removed from the Si wafer. The PL measurement on interface layer was done after the ASH layer has been removed from the wafer by rinsing the sample in DI water. The complete removal of the ASH layer from the sample was confirmed by FTIR measurements through the absence of N-H vibrations of $(NH_4)_2SiF_6$ species as demonstrated in Figure 1. The N-H vibrational modes of these species at 3328 cm-1 and 1436 cm-1 are absent in DI-rinsed sample, thus confirming the complete removal of the ASH from the wafer.

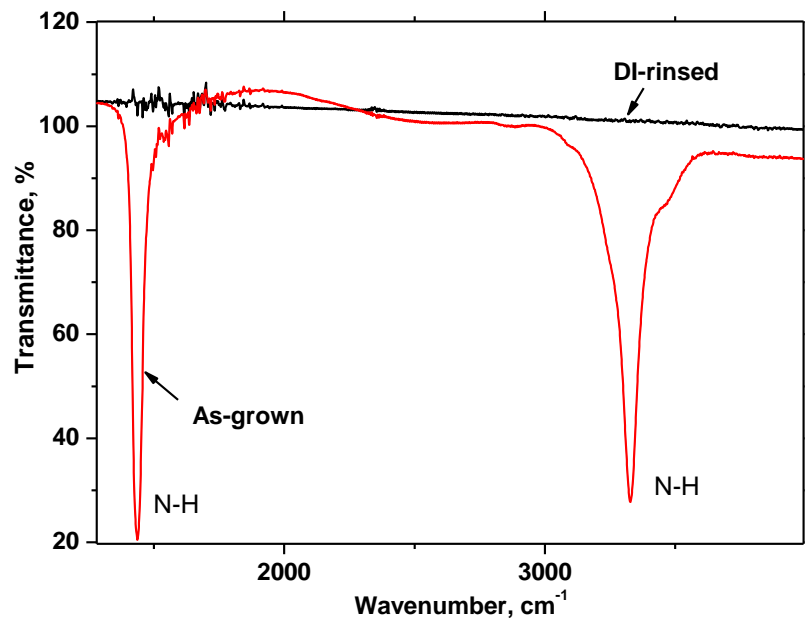

**Figure 1**

## 3. Results and discussion

Room temperature PL from the ASH/Si structure reveals a strong broad emission that is visible to naked eye in dark room. Figure 2 compares the PL spectra measured at 300 K for the whole as-grown sample (spectrum A) and for the same sample after the ASH layer has been removed (spectrum B) from the wafer. Actually, the spectrum B corresponds to an emission originating from a very thin porous oxide interface layer with a thickness of up to 500 nm depending on the duration of the exposure time. The difference between the A and B (that is the dominant peak at 605 nm, red line) is the major emission, which originates from the ASH layer (Fig. 2). The shoulder of the peak A at around 700 nm (1.77 eV) is a PL component associated with the ASH-Si wafer interface. Both A and B can be deconvoluted into three bands at 605 nm, 660 nm and 700 nm assuming a Gaussian curve fit. The presence of these bands at the same energies is suggestive of the incorporation of smaller size Si particles into the ASH layer (A) if one considers quantum confinement effects [12]. The presence of a high energy emission in both of the structures suggests that the emission A from the free standing ASH layer has the same origin as the emission from the interface layer. With the difference that the ASH layer contains greater number of such particles. In other words, the PL emission from the ASH layer can be associated with oxygen related defect in or around the Si nanoparticles embedded into the ASH during the transformation process of Si to $(NH_4)_2SiF_6$ crystals. The interface layer can be regarded as a kind of porous oxide(SiOx) layer, comprising of Si and oxygen as evidenced from the EDS and FTIR measurements [7]. Therefore, we suggests that Si and silicon oxide species originating from the interface are embedded into the ASH matrix during the transformation process. However, the size of Si particles embedded in ASH must be much smaller than those incorporated in the interfacial layer due to the much stronger intensities of the high energy bands ($\lambda \leq 605$ nm).

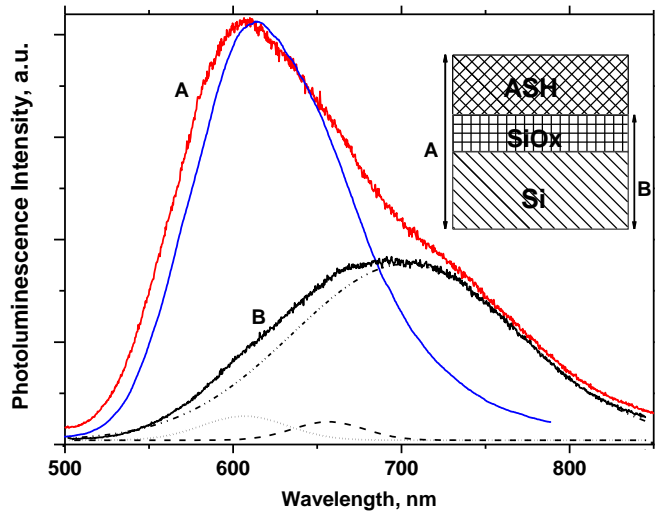

**Figure 2**

Depending on the theoretical and experimental results reported in litterature taking into account quantum confinement effects, the energetic position of the strongest band (2.05 eV) in our material would correspond to an average size of the Si NPs to be between 1-3 nm [12 and references therein]. The same peak energy would correspond to silicon clusters of about 5 to 20 atoms in diameter depending on the estimation methods [12' and references therein]. Using the mass density of bulk silicon, one can readily calculate that 20 atoms fill a cluster of 9 Å in diameter. However, there are still questions about their nature: are these NPs crystalline or amorphous or are they Si-rich $SiO_x$ particles? The role and the nature of the possible defects are yet to be determined since physical properties are very sensitive to the number of atoms.

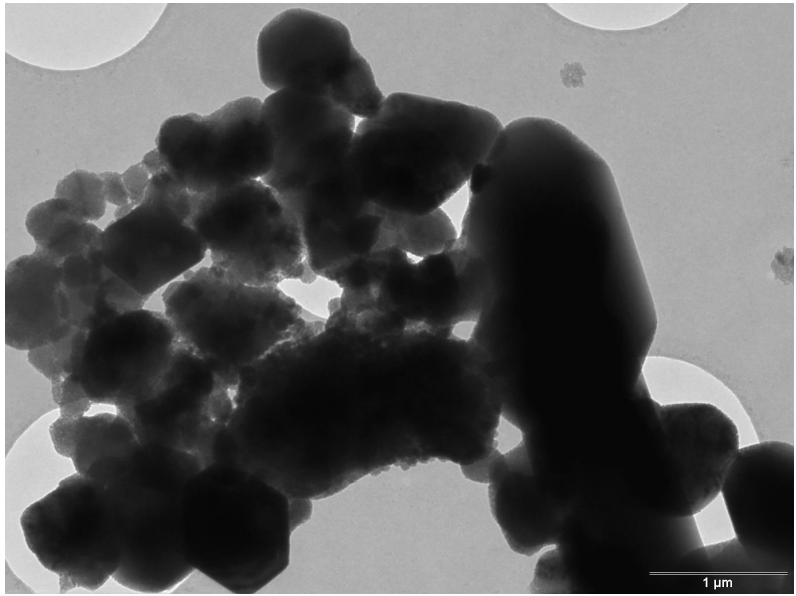

Figure 3a

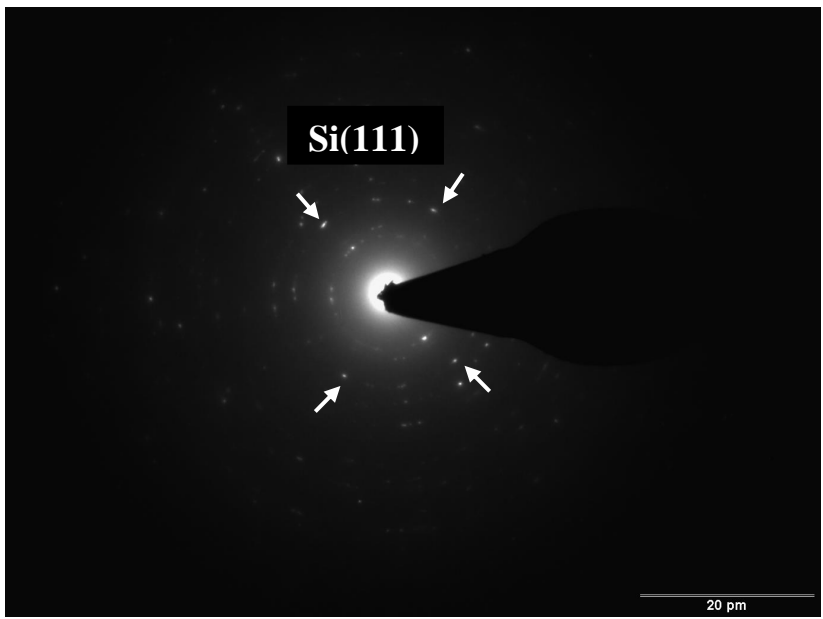

**Fig 3b**

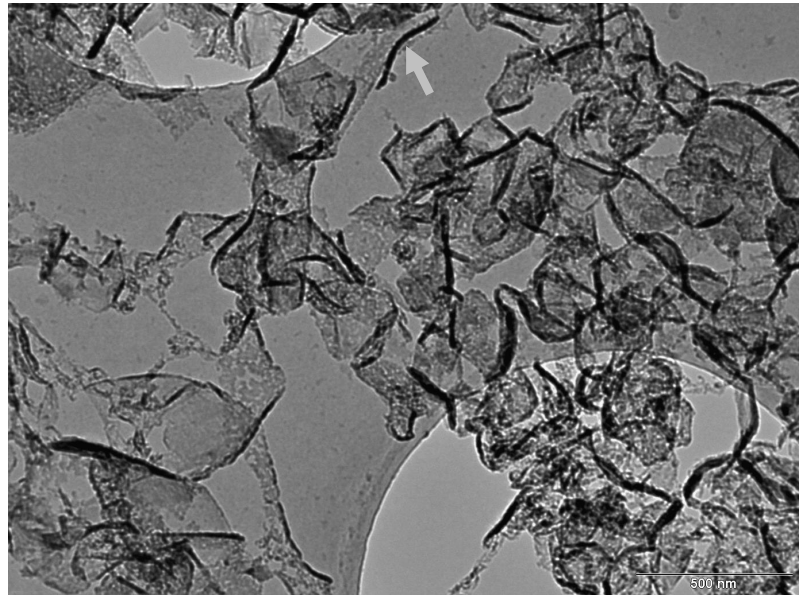

**Fig 3c**

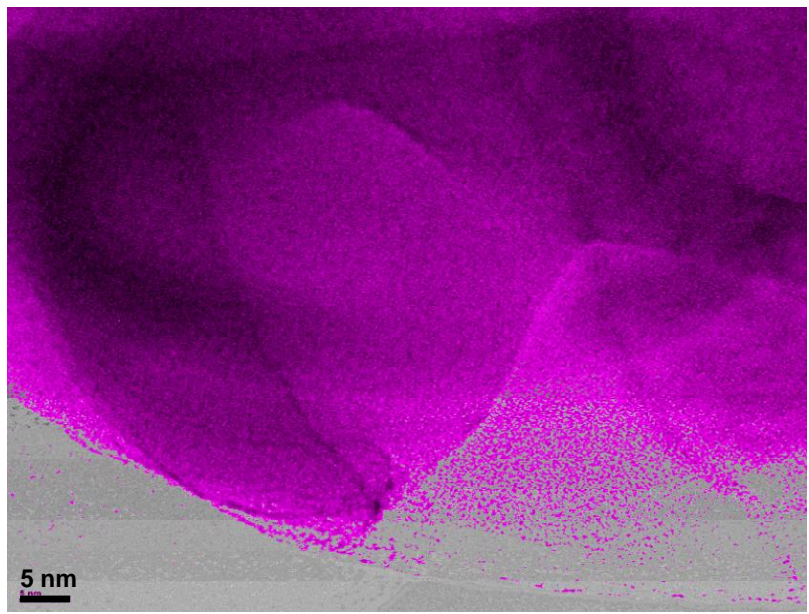

**Fig 3d**

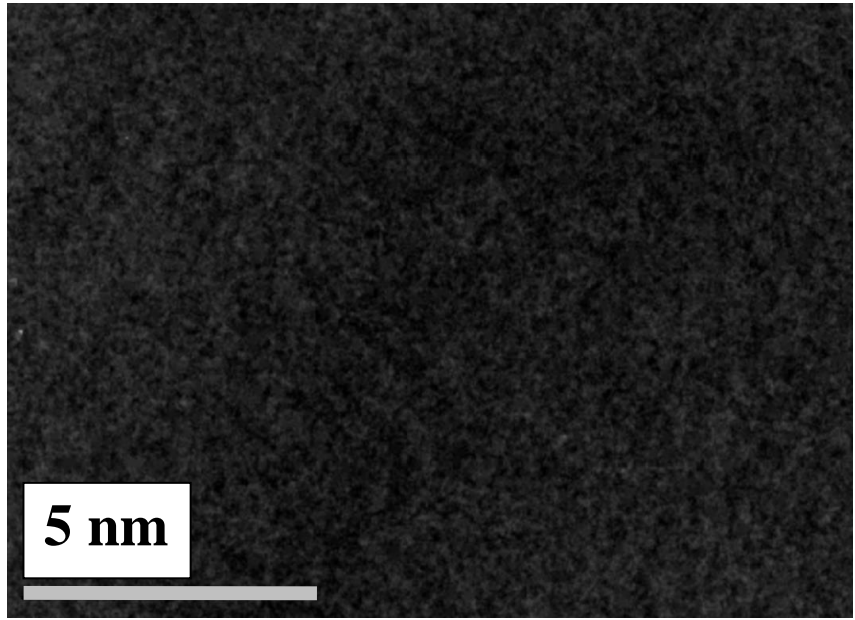

**Fig 3^e**

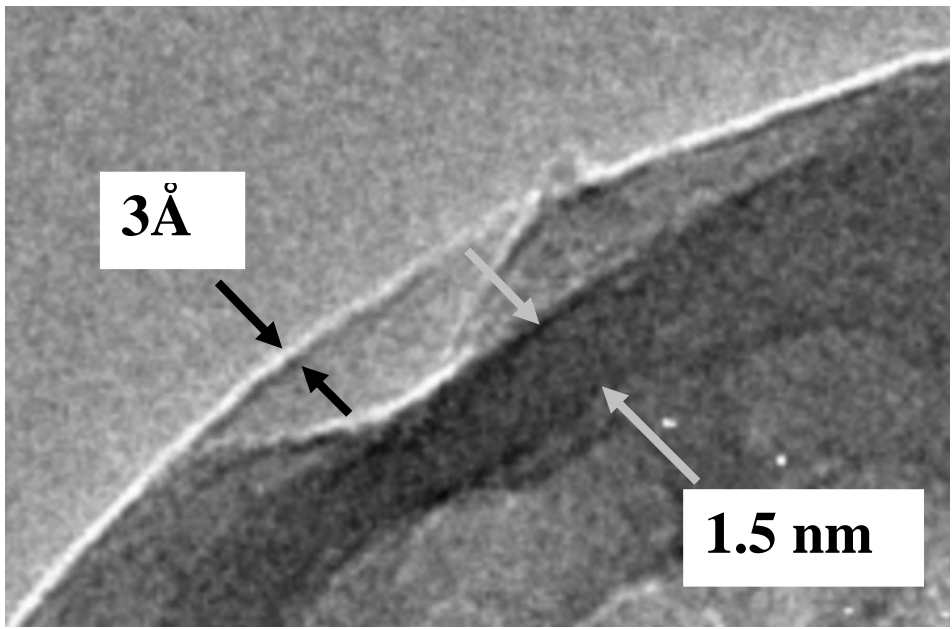

Indeed, TEM image in Fig. 3a, as taken from a free standing ASH layer, confirms the presence of sub-micron size clusters having a typical hexagonal crystal structure of $(NH_4)_2SiF_6$ [4]. These large clusters are distinguished from the surrounding by their dark color due to the presence of silicon particles which are embedded within these crystallites. Diffraction pattern taken under TEM also confirms the presence of a crystalline structure in Fig. 3b. The diffraction patterns can be interpreted as a mixed crystals of Si(111) and $(NH_4)_2SiF_6$. Unfortunately, the instability of the ASH material under e-beam exposure limits the observation time and consequently does not allow large magnification for lattice imaging. Subsequent TEM imaging of the same structure, as shown in Figure 3c shows the slabs and wire-like features formed by the Si particles as a result of the collapsed large ASH crystals. Since the ASH crystals are not thermally stable, an exposure to e-beam decomposes these crystals leading to the formation of the slabs and wire-like structures. Figure 3c indeed represents the crystals which have been exposed to e-beam for a few seconds. The image indicates the presence of a phase containing much smaller particles (less than few nanometers) with lower contrast throughout the layer with the presence of wire or ribbon-like features (white arrow) around the peripheral areas of the clusters. Previous studies indicated that the ASH has a polycrystalline structure [4]. Thus, it is reasonable to assume that the Si-NPs are rather located on the surfaces of these crystals.

In Figure 3d, TEM image reveals that most of the particles are connected between each other, thus leading to a continuous morphology. Large magnification of the same image (Figure 3e) actually represents that a porous-like structure exists as it was also supported by the absence of diffraction pattern. However, the layer appears rather discontinuous in the form of a dispersion of grains having particles sizes of less than 2 nm. The dark regions could be attributed to a local increase of the thickness due to stacking of particles. The absence of lattice images could be due to a natural disappearance of the crystalline contrast for silicon particles of less than 2nm as reported earlier [12''goldstein] [12'''hofmann]. The formation of an oxide layer of oxide around Si particles and even a possible oxygen diffusion could further complicate the observation of lattice fringes.

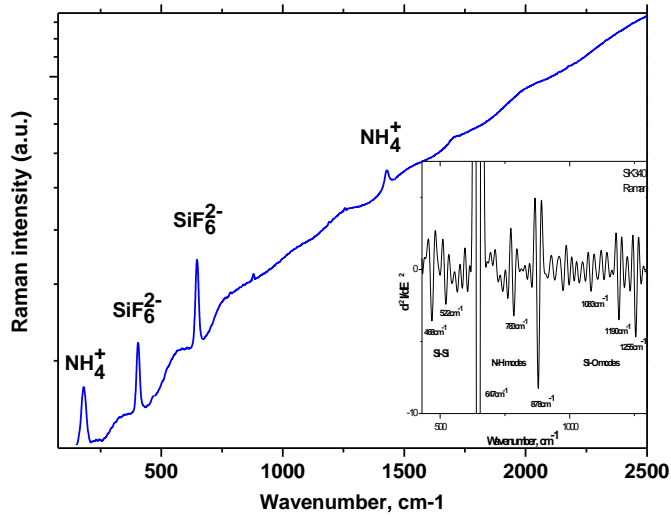

**Figure 4**

The peaks observed in a typical Raman spectrum of the ASH layer is presented in Figure 4. The major strong Raman peaks observed at 181 cm$^{-1}$, 402 cm$^{-1}$, 645 cm$^{-1}$ and 1427 cm$^{-1}$ are the lines associated with the vibrations of $NH_4$ and $SiF_6$ [Poulet 1976]. Due to the background luminescence signal, the remaining lines are barely observed. In order to distinguish these lines, a second derivative of the PL signal is inserted in the same figure. Si-O-Si stretching modes can be distinguished at 1083, 1190 and 1255 cm-1. We don't observe the usual Si-Si TO phonon line at 522 cm-1 but the band at around 468 cm-1 is attributable to TO phonon line of Si nanoparticles [12'''']. From this band, particle size was estimated to be about about L≈2.0 nm using the following expression [12'''''],

$\Delta\omega=\omega(L)-\omega_o=-A(a/L)^\gamma$

**Where ω, L, A and a are the line shift, particle size,**

In Raman we dont clearly observe TO phonon line, probably due to the photoluminescence background and size related quenching effect and redshift of the phonon line. A red-shift and quenching in intensity of the 1st and 2nd order Raman peaks (TO phonon at **Γ** and **L** points) with decreasing size has already been reported by several groups [32].

Concerning the origin of the observed PL bands, they would correspond to crystal sizes of about 1.0 to 3.0 nm [23, 24, 24'] if the quantum confinement is in effect. TEM analysis suggest that these structures are rather amorphous, but oxydized to some degree as evidenced by the presence of Si-O bonds in FTIR measurements on free standing ASH. The EDS analysis support the presence of oxygen of up to 4.5 at.% in ASH. Furthermore, TEM diffraction patterns yielded that in some of the samples of the ASH layers do not show any crystalline content (see Fig. 3d) while emitting visible light. This observation rather suggests that the amorphous phase is dominant or Si particle sizes are smaller than the TEM resolution. Another feature of these nanostructures is the encapsulation by SiOx ($1 < \mathbf{x} < 2$). In both cases, the PL should be excited within the confined structures surrounded by oxygen-rich SiOx. The recombination might occur in the same structures via defect states at interfaces [10]. This suggests that both the confinement effect and defect states are involved in the radiative recombination process.

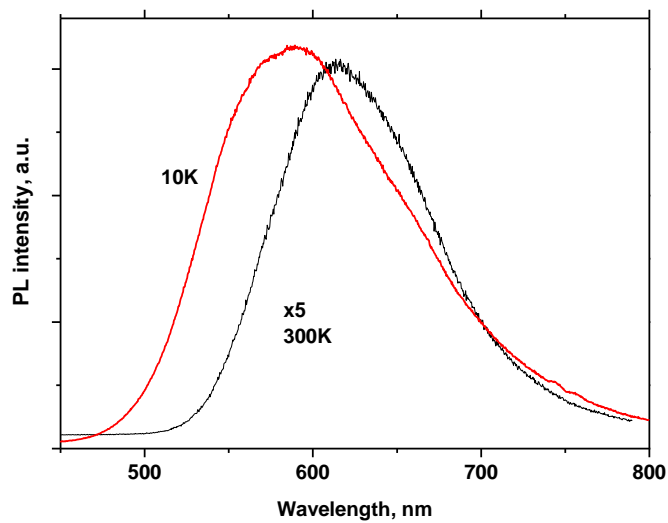

**Figure 4**

Temperature dependent PL measurements provide a further insight to the nature of the PL mechanism. The low temperature PL at 10K has a much stronger intensity and a blue shift of 186 meV at the high energy edge of the emission as seen in Fig. 4, which was followed by a peak broadening from 350 meV to 490 meV. This change can be attributed to quantum confinement effect in Si nanoparticles but also to the spatial confinement as described earlier [25-27]. We assume an oxygen diffusion into Si core resulting in a Si rich SiOx core and oxygen-rich SiOx shell and a composition gradient between the core and the shell structure. This effect would result in a Si-SiOx confined structures as well as in a corresponding band structure as sketched as insert in Fig. 4. The PL can be excited within the Si or Si-rich Si-SiOx and in the surrounding SiOx layer. As the temperature raises the carriers in shallow potential wells are thermally transferred to deeper potential wells and thus leading to an increased PL intensity at lower energies (Fig. 4 at 300K). In both cases of the confined structures, the recombination most probably takes place through defect states at Si-SiOx interfaces or in SiOx. Therefore such oxygen related defects whose, origin is not yet known determine the PL properties. Regardless of sample preparation conditions, the major peaks were observed in all the samples at about the same wavelengths: 550 nm (2.26 eV), 570 nm (2.18 eV), 605 nm (2.05 eV), 665 nm (1.86 eV) and 740 nm (1.68 eV). The difference lies only at the relative band intensities depending on the preparation conditions. The defect origin of the emissions is also supported by low temperature PL wherein almost no significant shift at the emission energies was observed as shown in the second derivative PL spectrum Fig.4. Relative insensitivity of these bands to temperature indicates that recombination between localized states are involved [25].

The blue shift of 186 meV at the high energy edge of the emission in Fig. 4a is attributable to a more efficient carrier trapping in smaller particles (Si or oxygen-rich SiOx) at low temperatures. As the temperature raises carriers are trapped in O-rich SiOx and diffused out to Si-rich SiOx or larger Si particles at high temperatures The realization of this event can be understood by the following arguments: In highly confined amorphous Si the band-gap energy change of up to 200 meV can be observed [26]. The bandgap change in such structures could be explained by spatial confinement of carriers recombining via localized tail states. Any lattice dilation effect can be neglected due to smaller thermal expansion and compressibility in Si nanostructures[27]. Then the critical distance for spatial confinement [26] can be estimated to be about 20 A and 500 A at 10K and 300K, respectively. TEM analysis indicates there is enough number of particles within the critical volume defined by these radiuses. Thus, one can assume similar charge transfer effects between in Si-NPs (O-rich SiOx and Si-rich SiOx) but the tunneling is made from the quantum well to defect states and the subsequent recombination occur through the defect states instead of tail states. Therefore, the increased intensity of the emissions at around 550 nm should be the main reason for the blue shift at the PL edge.

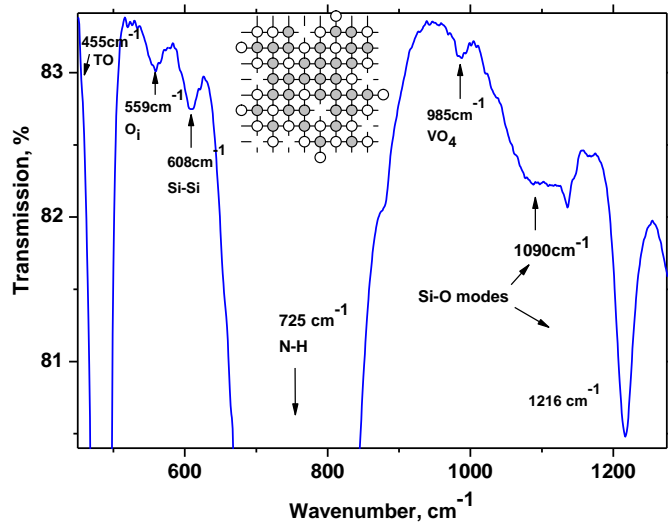

**Figure 5**

Increasing absorption toward high energy in Si-O-Si stretching modes is suggestive of the presence of oxygen-rich SiOx with **x** much larger than 1.0 [27'].

In addition to the TO phonon band in 1000-1200 cm-1 region, the sharp feature at 1216 cm-1 is attributable to the LO band of Si-O-Si vibrations. Such sharp LO band has already been observed in chemically formed SiO2 thin layers [27''].  These studies show that the peak position of the LO band increases depending on the oxide thickness. From the interpolation of this dependence, the peak position at 1216 cm-1 can be found to be corresponding to a SiOx layer thickness of 1.65 nm.

The stronger overall PL emission intensity in the ASH should be attributed to an increase in number of Si-NPs due to the layer thickness, which is at least 20 times thicker than the interface oxide layer. However, one can also argue that the ASH layer is an effective host for these radiative recombination centers. An efficient PL emission requires that the Silicon clusters are encapsulated or surface defects are passivated by oxygen, hydrogen, fluorine or nitrogen atoms [11] [28-29]. This is actually true if we take a closer look at the FTIR measurements on a free standing ASH layer revealing the presence of Si-O related vibrations at 1085 cm-1(AS1-TO1) and 1135 cm-1(LO) and 1216 cm-1 (AS2-TO2) as shown in Fig. 5a.  Hence it is not surprising that an oxydation layer of amorphous SiOx encapsulates Si particles.  In addition, the same spectrum shows the presence of Si-Si lattice vibrations at 608 cm-1 and Si-Si TO phonon vibration at 455 cm-1. Additional support regarding Si-Si and Si-O modes comes from the Raman scattering measurements on the ASH layer in Fig. 5b. The co-existence of 468cm-1(TO phonon) and 522cm-1 Si-Si vibration confirms the presence of amorphous and crystalline phases, thus supporting FTIR results. These vibrational modes are mostly supportive of the presence of silicon clusters which are encapsulated by silicon oxide.  Note that the large absorption band at 725 cm-1 is one of the main N-H bands of the ASH molecules[2, 4, 7]. Additional proof for oxide encapsulation comes from the EDS measurements confirming also the presence of oxygen within the layer as evidenced from the cross-sectional analysis of the ASH structure. The results on a number of samples reveal up to 4.5 at.% of oxygen within the ASH layer. Moreover,  FTIR spectrum taken on a free standing ASH as shown in Fig.5 reveals the presence of an interstitial oxygen $O_i$ [30-30'] as evidenced by absorption bands at 560 cm-1 and the band at around 1100 cm-1. Another possible type of defect could be a $VO_4$ complex of a vacancy with four oxygen atoms as evidenced by the absorption band at 985 cm-1[31].

The SE measurements were carried out in the wavelength range of 250-1000nm with a step of 5 nm and the incidence angle was set at 80o.

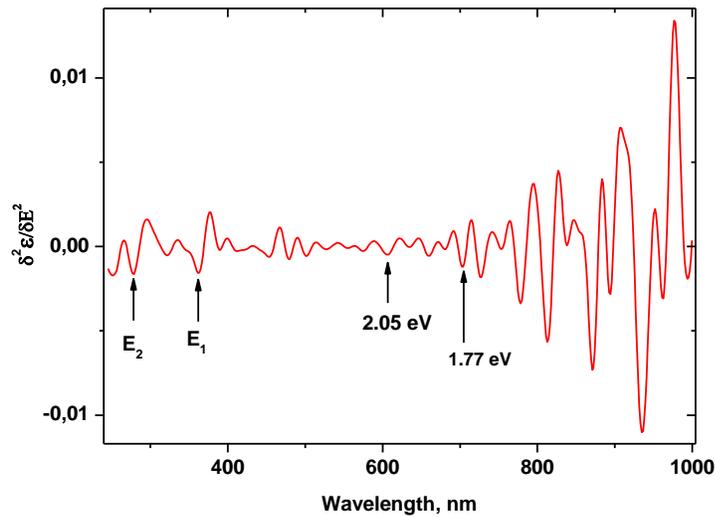

Figure 6 shows the second derivative of the dielectric function revealing the electronic critical point energies of our material. The increase in $E_1$ andc $E_2$ energies is indicative of a possible quantum confinement effect. There is also a peak at 605 nm corresponding to the most intensive peak observed in PL spectra. In lower energies, that is $\lambda > 700$ nm, the spectrum is difficult to interpret due to interference fringes.

No SiO2 peaks in XRD [4] thus → nonstoichiometric amorphous SiOx

However, the observation suggests that there should not be an abrupt interface between Si and SiOx. We propose a type of interfacial structure with a gradient in oxygen concentration between Si and SiOx. This type of structure should also involve vacancies and dangling bonds as well as indicated at the insert in Fig. 5a. Our previous work has shown that the interface between Si and SiOx is not sharp[2] revealing the diffusion of the oxygen to the Si core.

This effect can transform the core Si to a Si rich SiOx and as a consequence one can expect a direct emission from this type of a core structure. However, we can not rule out also the possibility of having carriers excited in Si NC's and/or oxygen rich SiOx and a recombination through defects on the Si surface, Si-SiOx interface and defect states in SiOx.

**4. Conclusion**

We have investigated the strong PL emission from the ammonium silicon hexafluoride dielectric layers obtained from the transformation of Si wafer surface treated under acid vapors. TEM and EDS measurements which have been combined with FTIR, Raman analysis show that the PL emission from the ASH layer is originated from confined Silicon nanoparticles within the ASH. These structures were incorporated into the growing layer during transformation of Silicon to $(NH_4)_2SiF_6$. The results show that there are both amorphous and crystalline phases of Si-NPs. They can be in the form of Si and Si-rich $SiO_x$ particles passivated by oxygen rich SiOx and are located on the surfaces of the bulk ASF polycrystals of sub-micron size. The PL can be either excited at confined Si-rich $SiO_x$, O-rich SiOx or in Si nanostructures but the recombination occurs through quantum confinement and localized defect states involving oxygen. The domination of the blue-shifted peaks as compared to those from the interfacial porous oxide layer suggests that smaller Si nanoparticles are embedded into the ASH layer.

Observation of similar group of photoluminescence bands regardless of sample preparation conditions suggests that rather defects determine light emission properties of the ASH/Si layers. The nature of these defects has been speculated through the observation of oxygen related vibrational modes.

The ASH layer could be an efficient host for light emitting Silicon particles. However, future work is required to understand the detailed microscopic role of this host in PL kinetics, type of defects and control the process in order to explore the possibility of fabricating a gain medium for luminescent devices.

**Acknowledgement**

This work was supported by BMBF-TUBITAK bilateral program under contract No:107T624, European FP6 Research Infrastructures program through MC2 ACCESS – contract No:026029 and German Federal Ministry of Education and Research (Grant No: 03Z2HN12).

[32] Meier C, Lüttjohann S, Kravets V.G, Nienhaus H, Lorke A and Wiggers H 2006 Physica E **32** 155

[  ] Poulet H and Mathieu JP 1976 J Raman Spectroscopy **5** 193

**FIGURE CAPTIONS**

**Figure 1.**  SPM image of the interfacial layer after the ASH layer has been removed from the Si wafer.  The image was taken on 1 μm x 1 μm area. The roughness is RMS=33.9nm with the cones of up to 60 nm high.

**Figure 2.**  Room temperature PL emission from the ASH/Si system. a) Light emission as compared to interface and b) Second derivative PL signal.  The peak A is originated from the ASH/Si layer and the peak B is from the porous oxide layer at the interface as indicated at the insert.

**Figure 3.**  a) HRTEM image of the ASH layer reveals the presence of  bulk ASH crystalline clusters and Si nanoparticles, b) x-ray diffraction pattern obtained from these clusters, c) a sketch indicating the formation of slabs and wires as a result of shrinking ASH crystals by e-beam induced heating.  Note that the prolonged exposure to e-beam enhances the transparency and leads to a clear image of  Si nanoparticle slabs and wire-like features (as shown by an arrow).

**Figure 4.**  Temperature dependence of photoluminescence from the ASH/Si (a) and its derivative (b) at 10K and 300K.  Note that there is a shift of 186 meV at the high energy side of the emission due to the presence of a greater number of smaller particles in ASH layer.  The inserted picture shows a schematic band structure for carrier relaxation and possible recombination pathways through defects in SiOx or Si/SiOx interface to explain the PL  from the Si clusters formed within the ASH matrix.  At low temperature carriers are trapped in O-rich SiOx and diffused out to Si-rich SiOx or larger Si particles at high temperatures.

**Figure 5.**  a) FTIR spectrum indicating the presence of TO phonons, interstitial oxygen $O_i$, Si-Si, VO4, Si-O and N-H major bands.  The insert indicates a possible Si particle structure wherein open circles represent oxygen atoms, b) Second derivative of the Raman scattering spectrum for the ASH on Si system indicating Si-Si, Si-O and N-H vibrations.